\documentclass[11pt,twoside]{article}

\def \etal {et~al.~}
\def \Vrot {\ifmmode V_{\rm rot} \else $V_{\rm rot}$ \fi} 
\def \Vvir {\ifmmode V_{\rm vir} \else $V_{\rm vir}$  \fi} 
\def \Vmax {\ifmmode V_{\rm max} \else $V_{\rm max}$  \fi} 
\def \Rd {\ifmmode R_{\rm  d} \else  $R_{\rm d}$  \fi} 

\usepackage{asp2006}
\usepackage{epsf}
\usepackage{psfig}
\usepackage{lscape}
\usepackage{graphicx}

\begin{document}
\title{Confronting   Scaling  Relations   of   Spiral  Galaxies   with
  Hierarchical Models of Disk Formation} 

\author{Aaron A. Dutton\altaffilmark{1} and St\'ephane Courteau\altaffilmark{2}}
\altaffiltext{1}{UCO/Lick Observatory and Department of Astronomy and Astrophysics, University of California, Santa Cruz, CA 95064}
\altaffiltext{2}{Department of Physics, Queen's University, Kingston, ON K7L 3N6 Canada}

\begin{abstract}
  The scaling relations between rotation velocity, size and luminosity
  form a benchmark  test for any theory of  disk galaxy formation.  We
  confront  recent theoretical models  of disk  formation to  a recent
  large  compilation  of  such   scaling  relations.   We  stress  the
  importance  of  achieving  a  fair  comparison  between  models  and
  observations.
\end{abstract}

\vspace{-0.5cm}
\section{Introduction}
Understanding the origin  and nature of galaxy scaling  relations is a
fundamental quest  of any successful theory of  galaxy formation.  The
success  of a  particular  theory will  be  judged by  its ability  to
predict  the  slope, scatter,  and  zero-point  of  any robust  galaxy
scaling relation at any  particular wavelength.  The scaling relations
between rotation velocity, $V$, size, $R$, and luminosity, $L$, are of
special interest as they are linked via the virial theorem.

Courteau \etal  (2007; hereafter C07)  compiled a sample of  1303 disk
galaxies  for which accurate  rotational velocities  and near-infrared
sizes and luminosities are available.   We refer the reader to C07 for
details about this compilation.  The $VLR$ scaling relations for these
galaxies are  shown in  Fig.~1.  The solid  black lines in  each panel
show orthogonal linear fits to the combined data set, while the dashed
lines show the  $2\sigma$ scatter in these relations.   The points are
color-coded according to central surface brightness.  This reveals the
fundamental independence  of surface brightness of  the $VL$ relation.
As discussed  in C07,  the $VL$ relation  is also independent  of disk
size  and  residuals  from  the  size-luminosity  relation.   This  is
unexpected, as the simplest models of galaxies embedded in dark matter
halos  predict a  strong  surface brightness  dependence, unless  disk
galaxies  are dominated by  dark matter  within their  optical regions
(Courteau \& Rix 1999; Dutton \etal 2007).

\section{Comparison with hierarchical galaxy formation models}
\begin{figure}[ht]
\begin{center}
\includegraphics[width=0.80\textwidth]{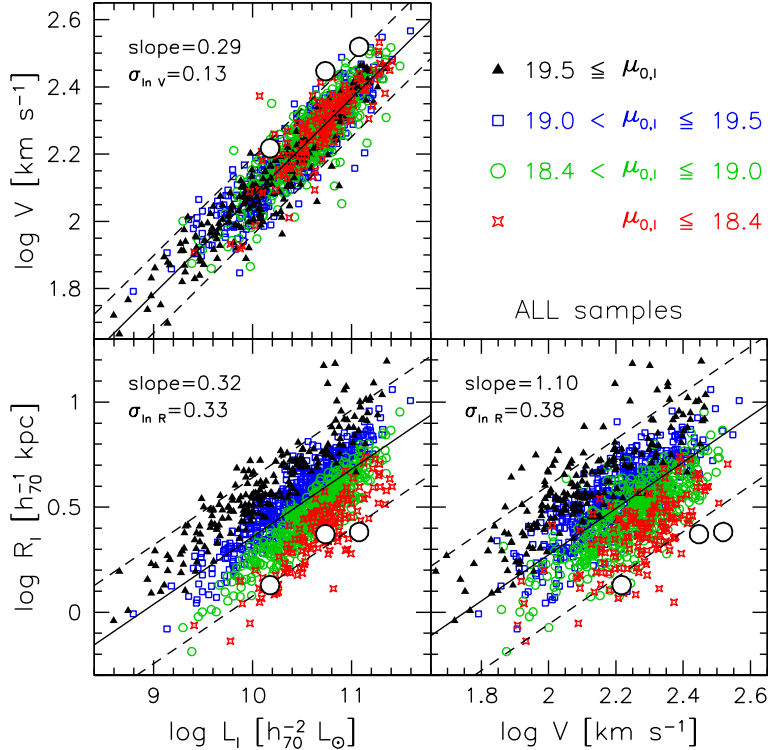}
\caption{VLR  scaling   relations  color  coded   by  central  surface
  brightness.   The size-luminosity  relation has  a  (trivial) strong
  surface   brightness   dependence,   but  the   surface   brightness
  independence  of the  velocity-luminosity  goes against  theoretical
  expectations. The white filled  circles show the models of Governato
  \etal 2007; see text for details.}
\label{fig:VLRmu}
\end{center}
\end{figure}
Cosmological  numerical  simulations  of  galaxies  have  so  far  all
encountered the  same fundamental  problem: model galaxies  rotate too
fast and are too compact  compared to observed ones (e.g. Eke, Navarro
\&  Steinmetz 2001;  hereafter ENS).   Whether this  is the  result of
numerical artifacts or a reflexion  of incomplete physics, or both, as
yet  to be  resolved.  However,  the recent  simulations  by Governato
\etal (2007;  hereafter G07) would  have, according to  their authors,
overcome   difficulties  of   yester  efforts   and  achieved   a  new
breakthrough.  Quoting from their abstract:

\vspace{0.2cm}
{\it ``The simulated  galaxies form rotationally supported disks
    with  realistic exponential  scale lengths  and fall  on  both the
    I-band and baryonic Tully Fisher relations.''}
\vspace{0.2cm}

However, as we discuss below, a fair comparison between the G07 models
and observed measurements casts doubts about these claims.

\subsection{On establishing a fair comparison between models and data}
The  $VL$ relation  has long  been a  benchmark test for hierarchical
models of galaxy  formation.  In order to establish  a fair comparison
between  observed and  predicted scaling  relations, all  the relevant
parameters  must be  measured  in  a similar  way  between models  and
observations  and drawn  from comparable  parent  samples.  Rotational
velocities  from resolved  rotation curves  are typically  measured at
2.2R$_d$, where R$_d$  is the disk scale length  (Courteau 1997).  The
total luminosity  and $R_d$ are  also typically measured  from near-IR
surface brightness  profiles which  probe the underlying  stellar mass
nearly free of extinction.

G07 measured  the rotation velocities  of their model galaxies  at 2.2
and 3.5 $R_d$.  Bright disk  galaxies have flat rotation curves beyond
2-3 $R_d$ and the specific  choice of velocities matters little.  This
would not be true for  fainter galaxies whose rotation curves may keep
rising up to the last data  point.  The situation is backwards for G07
whose rotation  curves for all  3 models show  a steep inner  rise and
rapid decline in the outer  regions.  Care must thus be exercised when
measuring a fiducial rotational velocity.

Furthermore,  G07 use  $B$-band  scale lengths  which  are, for  their
models, either twice as large or  twice as small as the $K$-band scale
lengths.   As a  result, G07  measured $V_{rot}$  either significantly
within or  beyond the actual  maximum rotation velocities for  their 3
model galaxies.  Observations however show that $B$-band scale lengths
are typically 20\% greater than those at $K$ band (de Jong 1996).  The
wide range of $B/K$ disk scale lengths in G07 is thus unexpected.

G07 state that {\it ``at the resolution of our standard runs the total
  mass  profile and  the quantity  of  stars have  converged at  radii
  corresponding  to a  couple  of disc  scale  lengths.''}.  Thus  the
rotation curves should be resolved at 2.2$R_d$ measured in the K-band.
From their  rotation curves we  can estimate $V_{2.2K}=$165,  280, and
330 km s$^{-1}$ for the 3 model galaxies.  Fig.~1 shows that these new
models  create galaxies  that  fall  more than  $2\sigma$  off the  VL
relation.  Fig.~1  also shows the RL  and RV relations,  where we have
used the  K-band sizes from G07.   Our data use  I-band scale lengths,
and so  to make a  fair comparison we  have increased the  model scale
lengths by  0.05 dex (de  Jong 1996).  Nevertheless, G07  galaxies are
too small by  a factor $\sim2$ for a given luminosity  and by a factor
$\sim2.5$ for  a given velocity, which again  corresponds to $2\sigma$
offsets from the mean relations.

Thus a fair comparison between  the models and data indicates that the
G07  model galaxies  suffer  the same  fate  as previous  cosmological
simulations (e.g. ENS),  namely that they rotate too  fast and are too
small. It  should be  noted however, that  ENS measured  velocities at
radii which  enclosed essentially all  of the cold baryons  at $r_{\rm
  gal} =  20 (V_{200}/220) h^{-1}  \rm kpc$.  For the  G07 simulations
these radii correspond to 9, 18 and 24 kpc.  Using rotation velocities
measured at  such large radii the  G07 galaxies do indeed  fall on the
I-band, stellar mass  and baryonic TF relations. Thus,  in this sense,
the G07  simulations are an  improvement over ENS.  However,  the main
challenge for galaxy formation models is to produce disk galaxies with
the  correct  distributions of  baryons  and  dark  matter within  the
central few  kpc.  While G07  claimed that their  simulations achieved
this landmark, our analysis suggests the opposite.

\subsection{Have the simulations converged?}
We have  so far dwelled  on establishing fair  data-model comparisons.
Perhaps of even greater relevance  is whether the simulations are well
enough resolved  to warrant such  an exercise. G07 claimed  that their
simulations have converged beyond one stellar disk scale length, which
should  be adequate  for the  extraction  of reliable  disk sizes  and
rotation  velocities.   G07  ran   one  of  their  models  (to  z=0.5)
increasing  the number  of  particles eight-fold,  up  to 4.2  million
particles.  Convergence  of the rotation curves between  the two model
galaxies differing only in the  number of particles is only reached at
$\sim5$ kpc, corresponding to $\sim3.4 R_d$. This raises concerns that
the disk  sizes and rotation velocities  at $2.2 R_d$  have indeed not
been adequately resolved.

It is  no surprise  that these simulations,  like most others  at this
writing,  should  have  resolution  problems.  Kaufmann  \etal  (2007)
showed  that  even with  a  million  gas  and dark  matter  particles,
numerical effects result in spurious angular momentum losses, and this
is for  the idealized case  of a galaxy  growing in an  isolated halo.
For a  galaxy and halo growing hierarchically,  even higher resolution
is  needed  to resolve  the  progenitor  galaxies  and halos  at  high
redshift.

\subsection{The VL relation in semi-analytic models}
Current  semi-analytical  models (SAMs)  of  galaxy  formation face  a
similar challenge as N-body realisations.  Foremost, SAMs also fail to
match the zero point of  the $VL$ relation and the luminosity function
of disk  galaxies simultaneously (see  Dutton \etal and  Somerville in
these proceedings for more discussion).

Early SAMs  did not  model galaxy rotation  curves and  simply equated
$\Vrot$ with  the virial velocity  of the halo, $\Vvir$.   Modern SAMs
are often  only marginally more sophisticated. For  example the GALICS
SAM (Hatton  \etal 2003) use an  ad hoc function to  relate $\Vrot$ to
$\Vvir$, while for the Millennium SAM, Croton \etal (2006) assume that
$\Vrot$ is  equal to the maximum  circular velocity of  the halo, {\it
  prior} to  galaxy formation,  which is only  10 to 20\%  larger than
$\Vvir$ for  typical halo  concentrations (e.g.  Bullock  \etal 2001).
The GALform SAMs of Benson \etal (2003) are most complete by including
the  effects of  dissipating baryons  and halo  contraction.  However,
these  authors   measure  $\Vrot$  at  the  disk   half  mass  radius,
corresponding to  1.68 $R_d$  for an exponential  disk.  Observational
studies do not use rotation  velocity measurements at such small radii
since  they  are  more  susceptible  to  effects  of  inclination  and
non-circular motions.

Matching the zero point of  the $VL$ relation has been a long-standing
problem for hierarchical galaxy formation models, which may call for a
departure from  the standard theory  of disk galaxy  formation (Dutton
\etal 2007), or even require a modification of CDM on small scales.

\acknowledgements 
AAD   acknowledges  financial  support   from  the   National  Science
Foundation  grant AST-0507483.   S.C.  acknowledges financial  support
from the National Science and Engineering Council of Canada.

\end{document}